\documentstyle[pre,aps,psfig,floats,twocolumn]{revtex}
\begin{document}

\draft
\twocolumn[\hsize\textwidth\columnwidth\hsize\csname 
@twocolumnfalse\endcsname

\title{Bid distributions of competing agents in simple models of auctions}
\author{R. D'Hulst and G.J. Rodgers}
\address{Rene.DHulst@brunel.ac.uk and G.J.Rodgers@brunel.ac.uk}
\address{Department of Mathematical Sciences, Brunel University}
\address{Uxbridge, Middlesex, UB8 3PH, UK}

\maketitle
\begin{abstract}
Models of auctions or tendering processes are introduced. In every round of bidding the players select their bid from a probability distribution and whenever a bid is unsuccessful, it is discarded and replaced. For simple models, the probability distributions evolve to a stationary power law with the exponent dependent only on the number of players. For most situations, the system converges towards a state where all players are identical. A number of variations of this model are introduced and the application of these models to the dynamics of market makers is discussed. The effect of price uncertainty on bid distributions is presented. An underlying market structure generates heterogenous agents which do not have power law bid distribution in general. 
\end{abstract}

]
\narrowtext

\section{Introduction}
\label{sec:introduction}

A considerable part of the statistical physics community is interested in financial market mechanisms and related problems \cite{bouchaud,stanley}. One major challenge in this area is to give a detailed picture of the emergence of group quantities from market microstructure. For instance, how prices and their fluctuations are related to the balance of buyers and sellers. The major difference between this approach and that using traditional financial theories is that the emphasis is now put on comparing hypotheses and their implications to real market data. In this work, we investigate simple models of auctions, with various settings, and obtain explicit expressions for distributions that are intrinsic properties of sellers. As such, we are not able to provide any comparison with any measurable data, and in fact, from the simplicity of our models, it is unlikely that any convincing similarity could be spotted. Instead we present our work as a first step in auction modelling from a physicist's point of view, a problem that has not attracted the attention it deserves up to now. 

Much of economic activity is based on mini auctions or tenders in which a potential buyer offers a particular amount of money for a product or a potential seller offers to sell the product at a particular price. The recipient of this offer then compares it with the offers of competitor buyers or sellers to determine the best deal and consequently with whom to trade. Sellers who overprice their goods or buyers who are not prepared to spend sufficient money seldom trade and risk going out of business. There have been numerous works on auctions, and a useful summary can be found in Ref. \cite{ohara} and its references, while Ref. \cite{cohen} still provides a very interesting review of the subject. Much of this work has concentrated on modelling the generation of an equilibrium price, determined by some extremization procedure, either by profit maximization, by risk minimization, by considering inventory constraints, or by considering the price of transactions \cite{ohara}. These models are mainly concerned with the dynamics of price formation. In contrast, here we want to consider sellers competing to attract buyers, reducing their behaviour to a trial and error process. We want to model the learning of sellers that are repetitively competing against each other. In practice, our models are not specially devised to reproduce financial markets but rather to tackle the more general problem of competing sellers acting inductively \cite{conlisk}.

In an attempt to model this type of process we introduce simple models in which two or more players repeatedly bid against one another. Each player has a probability distribution from which they draw their bids at random. When a player is unsuccessful he discards that bid and replaces it with another bid selected at random. In practice, we do not associate the bid proposed by a player to a market price, but rather to the profit made by a player over a fair market price. As such, we let bids be in the range $(0,1)$, with 0 for no profit and 1 for a maximum profit. This implies that a bid is a simultaneous proxy for both the profit of a player and his risk. In the next section we introduce the two player system and solve it analytically in two particular cases; when both players have the same set of bids at the beginning and in the long time limit. We argue that, except for very specific situations, the system converges towards a symmetric situation for the players. In Sec. \ref{sec:the d player game} we solve two different $d$ player versions of the same game and in Sec. \ref{sec:market makers}, we extend the model to mimic market makers. We investigate in Sec. \ref{sec:price volatility as a measure of risk} the effect of price volatility on the bid distribution. In Sec. \ref{sec:implement a market structure}, we let players be heterogeneous by implementing a market structure, and solve exactly one simple situation. Our results are summarized in the last section, where we also discuss improvements to make the models more realistic.

In this work, we restrict our attention to the random picking of new bids from a uniform distribution. This minimalist adaptation process assumes that players do not have a very efficient record of past bids. But this is in line with the idea of players trying to make the maximum profit, while minimizing risk. If players only try to minimize their exposure, they keep track of the winning bids and no room is left for profit. By always picking bids from a uniform distribution, players keep trying to improve their profit. We will discuss on extending the models to incorporate more general adaptation processes, but we keep a general analysis of this problem for future work. 

We have to mention that we use the term auction to refer to the competition between buyers or sellers, but this does not compare with the usual definitions of auctions in the economics literature. Our auctions are concerned with the dynamics of intermediaries, trying to make a profit from the competitive sale of a commodity or a service. This is completely different from auction as defined in finance, where participants take part in several rounds of bids before a sale takes place.   
\section{The Two player game}
\label{sec:the two player game}

Imagine two players who each have an infinite set of numbers described by a probability distribution. At each time set the two players draw a number at random from their respective distributions. They compare numbers; the player with the smallest number wins and does nothing, the player who loses replaces his losing number in the probability distribution with another number chosen at random from a uniform distribution. We will call the players $P$ and $Q$ and their corresponding probability distributions at time $t$, $P(x,t)$ and $Q(x,t)$. The probability distributions obey the non-linear coupled integro-differential equations 

\begin{eqnarray}
\nonumber
\frac{\partial P (x,t)}{\partial t} &=& - P (x,t) \int_0^x Q (y,t) dy \\
&+& \int_0^1 P (y,t) \int_0^{y} Q (z,t) dz dy
\label{eq:evolution p,2 players}
\end{eqnarray}
and

\begin{eqnarray}
\nonumber
\frac{\partial Q (x,t)}{\partial t} &=& - Q (x,t) \int_0^x P (y,t) dy \\
&+& \int_0^1 Q (y,t) \int_0^{y} P (z,t) dz dy.
\label{eq:evolution q,2 players}
\end{eqnarray}
The first term on the right hand side in Eq. (\ref{eq:evolution p,2 players}) corresponds to the destruction of numbers in $P(x,t)$ when player $P$ draws a number larger than that drawn by player $Q$. The second term on the right hand side corresponds to the creation of new numbers in $P(x,t)$ after $P$ has lost. Eq. (\ref{eq:evolution q,2 players}) has similar terms. Providing that the initial distributions $P(x,0)$ and $Q(x,0)$ are normalised then we have

\begin{equation}
\int_0^1 P (x,t) dx = \int_0^1 Q (x,t) dx = 1
\end{equation}
for all time. We will find it useful to define the probability that $Q$ will win at time $t$, $\alpha  (t)$, by

\begin{equation}
\alpha (t) = \int_0^1 P (y,t) \int_0^y Q (z,t) dz dy
\label{eq:definition of alpha}
\end{equation}
and similarly the probability that $P$ will win at time $t$ by $\beta  (t) = 1- \alpha  (t)$. $\alpha  (t)$ and $\beta  (t)$ are the second terms on the right hand sides of Eqs (\ref{eq:evolution p,2 players}) and (\ref{eq:evolution q,2 players}) respectively.

We can solve (\ref{eq:evolution p,2 players}) and (\ref{eq:evolution q,2 players}) completely if $P(x,0) = Q(x,0)$. Then we have $P(x,t) = Q(x,t)$ for all time and

\begin{equation}
\frac{\partial P (x,t)}{\partial t} = - P (x,t) \int_0^x P (y,t) dy + \frac{1}{2}.
\label{eq:evolution p,same initial conditions}
\end{equation}
Introducing the cumulative probability distribution

\begin{equation}
F (x,t) = \int_0^x P (y,t) dy,
\end{equation}
we can rewrite (\ref{eq:evolution p,same initial conditions}) in terms of $F(x,t)$ as

\begin{equation}
\frac{\partial F (x,t)}{\partial t} = - \frac{F^2 (x,t)}{2} + \frac{x}{2}.
\end{equation}
This is easily solved to give 

\begin{equation}
F (x,t) = \sqrt{x} \left( \frac{F (x,0) + \sqrt{x} + (F (x,0) - \sqrt{x}) e^{-\sqrt{x} t}}{F (x,0) + \sqrt{x} - (F (x,0) - \sqrt{x}) e^{-\sqrt{x} t}}\right).
\label{eq:f for the two player simple game}
\end{equation}
Consequently, for all initial conditions $P(x,0) = Q(x,0)$, the long time state is stationary with $F(x, \infty ) =  \sqrt{x}$, or $P(x,\infty  ) = 1/(2 \sqrt{ x})$.

We cannot solve (\ref{eq:evolution p,2 players}) and (\ref{eq:evolution q,2 players}) for general initial conditions, except in the long time stationary limit. This can be done by setting the derivatives on the left hand side of (\ref{eq:evolution p,2 players}) and (\ref{eq:evolution q,2 players}) to zero and dropping the time dependence. This reveals

\begin{equation}
P (x) = \alpha x^{\alpha - 1} \qquad \hbox{and} \qquad Q (x) = (1- \alpha ) x^{-\alpha}
\end{equation}
where

\begin{equation}
\alpha = \lim_{t\rightarrow \infty} \alpha (t).
\label{eq:long time limit of alpha}
\end{equation}
Consequently, for all initial conditions the stationary state is a one parameter family of power laws with the exponents equal to the negative of the probability that a player will win in the long time limit. This probability is itself determined by the initial conditions. From a symmetry principle, one expects the stable states to be extrema of a characteristic function of both $P (x)$ and $Q (x)$. It seems reasonable to expect that $\alpha$ and $1-\alpha$ will characterize these distributions, respectively. Forming simple functions from these two expressions gives three characteristic values for $\alpha$, namely, $\alpha = 0$, 1/2 and 1. We have performed a number of simulations to confirm that $\alpha = 1/2$ gives the stable solution for most initial conditions. In fact, with very specific initial conditions, the system also converges towards $\alpha = 0$ or $\alpha =1$, starting with these values as initial conditions, for instance. In practice, one would not except any of these peculiar conditions to be realised. If the model can mimic a real situation, $\alpha = 1/2$ is the only value that one should encounter. In other words, for most initial conditions, the system is driven towards a state where the two players are identical.

Of particular interest is the stationary distribution of the prices, which is equal to 

\begin{equation}
Z (x) = P (x) \int_x^1 Q (y) dy + Q (x) \int_x^1 P (y) dy
\end{equation}
or

\begin{equation}
Z (x) = \alpha x^{\alpha - 1} + (1- \alpha) x^{-\alpha} - 1.
\end{equation}
In the most common situation, that is, when $\alpha = 1/2$, $Z (x) = x^{-1/2} -1$. The moments of the price distribution are equal to

\begin{eqnarray}
\label{eq:definition of the moments}
M_n &\equiv& \int_0^1 dx x^n Z (x)\\
&=& \frac{\alpha (1-\alpha) - n (n+1)}{(\alpha +n)(1-\alpha +n) (n+1)}.
\end{eqnarray} 
In particular, the average price $M_1$ is given by 

\begin{equation}
M_1 = \frac{3\alpha (1-\alpha)}{2(\alpha + 1) (2-\alpha)}.   
\end{equation}
It achieves its maximum value for $\alpha = 1/2$. Hence, the adaptative process, even if very simple, is efficient because, of all solutions, the system selects the one that gives the sellers the maximum profit.

Note that the basic adaptation process can be improved. The model can be generalised so that when a player loses the new number received is drawn from a probability distribution $\omega (x)$ rather than from the uniform distribution. In this case, in the long time limit, we have

\begin{equation}
\begin{array}{cc}
P (x) = \alpha \omega (x) \left( \int_0^x \omega (y) dy \right)^{\alpha - 1}\\
Q (x) = (1 -\alpha) \omega (x) \left( \int_0^x \omega (y) dy \right)^{-\alpha}\\
\end{array}
\end{equation}
where  $\alpha$ is given by (\ref{eq:long time limit of alpha}). Again, $\alpha = 1/2$ gives the stable solution. The previous equations clearly shows that the power law found for the player distributions are not characteristic of competition, as any other distribution would give another result. This conclusion, namely, that the power law distributions are not robust, will be attained in several occasions in this work. 
\section{The \textnormal{$d$} player game}
\label{sec:the d player game}

We can obtain similar solutions for the $d$ player game where only the player with the highest number changes his distribution. In particular, when all the players have the same starting conditions they all have the same probability distribution at time $t$. This obeys the non-linear integro-differential equation

\begin{equation}
\frac{\partial P (x,t)}{\partial t} = - P (x,t) \left( \int_0^x P (y,t) dy \right)^{d-1} + \frac{1}{d}.
\end{equation}		
In the long time limit this evolves to the distribution 

\begin{equation}
P (x) = \frac{1}{d} x^{\frac{1}{d}-1}.
\end{equation}
When the initial conditions are unequal the probability distribution of player $i$ ($i$ = 1,..., $d$) evolves to

\begin{equation}
P_i (x) = ( 1 - \nabla_i ) x^{-\nabla_i} 
\label{eq: general d, distribution}
\end{equation}
where  $\nabla_i$ is the probability that player $i$ wins in the long time limit and 

\begin{equation}
\sum_{i=1}^d \nabla_i = 1.
\label{eq:sumtoone of the exponents}
\end{equation}
The same symmetry argument as in the previous section can be put forward here, and it suggests that $\nabla_i = 1/d$ is the stable solution for most initial conditions. The function we extremize is formed by the product of all the prefactors in Eq. (\ref{eq: general d, distribution}). This function appears naturally as a scale factor in the evolution equation, because all probability distributions are multiplied by one another. For instance, in Eq. (\ref{eq:evolution p,2 players}), we have to multiply $P$ and $Q$ in both terms on the right hand side. We checked numerically that $\nabla_i = 1/d$ characterizes the most common stationary state.

A more rigorous argument can be put forward by considering the price distribution $Z(x)$, defined as the probability that the selling price is equal to $x$. This distribution is given by

\begin{equation}
Z (x) = \left( \sum_{i=1}^d \frac{1 - \nabla_i}{x^{\nabla_i}-x} \right) \left( \prod_{i=1}^d (1 - x^{1-\nabla_i}) \right).
\end{equation}
The first moment of this distribution is the average price and is a function of the set of exponents $\lbrace \nabla_i; i=1,..,d\rbrace$. By looking to the extrema of $Z (x)$ with respect to these exponents, with the added condition Eq. (\ref{eq:sumtoone of the exponents}), one obtains that $\nabla_i = 1-1/d$ for all $i$ corresponds to a maximum of $Z (x)$. By extension, it is also a maximum of its first moment, meaning that the system reaches a stationary state where the players maximize their profit. In particular,

\begin{equation}
Z (x) = \left( \frac{1}{x^{1/d}} - 1 \right)^{d-1}
\end{equation}
when all players are identical. The moments $M_n (d)$ of this distribution, defined in Eq. (\ref{eq:definition of the moments}),  are given by

\begin{equation}
M_n (d) = \frac{d! (nd)!}{((n+1)d)!}.
\end{equation}
The average price,equal to $M_1 (d)$, is a decreasing function of the number of players. As this price compares with a profit, we can associate it to a measure of the spread between ask and bid prices. In this case, we conclude that the spread is a decreasing function of the number of players. This is a well-known fact that has been observed empirically.

As in the two player game, the previous model can be generalised so that when a player loses the new number received is drawn from a probability distribution $\omega (x)$ rather than from a uniform distribution. In this case, 

\begin{equation}
P_i (x) = (1-\nabla_i) \omega (x) \left( \int_0^x \omega (y) dy \right)^{-\nabla_i}
\end{equation}
in the stationary limit, which provides a slight improvement to the adaptation process.

In the previous model, only one player updates his distribution at each round of bidding. However, for auctions where agents are quoting prices to a buyer, the asset will only be sold by the agent proposing the lowest price. We consider the $d$ player game where all players that have not proposed the lowest price discard their proposal and draw a new price from a flat distribution. The players distributions follow coupled differential equations and for the particular case of all players starting from the same probability distribution, they all keep the same probability distribution during the whole game. This probability distribution obeys

\begin{eqnarray}
\nonumber
\frac{\partial P (x,t)}{\partial t} &=& - P (x,t) \left( 1 - \left( 1 - \int_0^x P (y,t) dy \right)^{d-1} \right)\\
&+& \frac{d-1}{d}
\label{eq:d player distribution one winner}
\end{eqnarray} 
where the first term on the right hand side shows that a player proposing $x$ is discarding this value as soon as another player proposes a lower bid. The second term means that a player is always changing his bid unless he wins. We have set equal to $1/d$ the winning probability of all players, because of the model symmetry. Note that from now on, we always consider symmetric initial conditions, because as we showed, this is the most common stationary state.

In the long time limit, Eq. (\ref{eq:d player distribution one winner}) is equivalent to the following equation

\begin{equation}
f (x)^d - f(x) d + (1-x)(d-1) = 0
\label{eq:f(x)}
\end{equation}
where we defined

\begin{equation}
f (x) \equiv \int_x^1 P (y) dy.
\end{equation}
It is easy to check that for $d=2$, the solution of Sec. \ref{sec:the two player game} is recovered. 

\begin{figure}
\centerline{\psfig{file=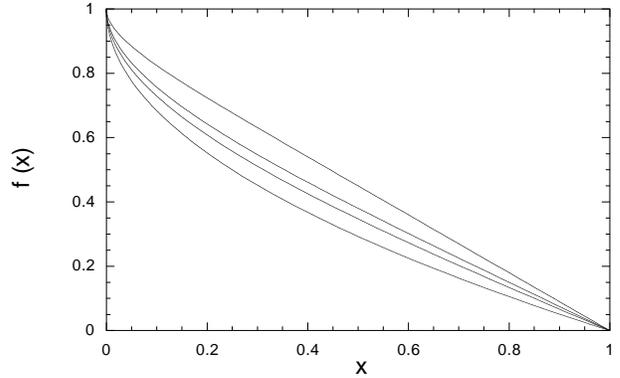,width=8.5cm}}
\caption{Cumulative distribution $f (x)$ for $d=2$, 3, 4 and 10, from bottom to top. As $d$ increases, $f (x)$ converges towards $1-x$. The solution is graphical as we only have $x$ as a function of $f$.}
\label{fig:f(x)}
\end{figure}

\begin{figure}
\centerline{\psfig{file=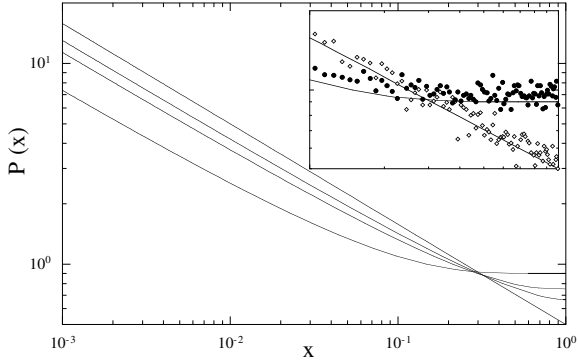,width=8.5cm}}
\caption{Probability distribution $P (x)$ for the bids of the agents for a game involving $d=2$,3,4 and 10 agents, from top to bottom on the left hand side of the figure. In the inset is presented a zoom of the region $x=0.1$ to 1. The continuous lines are analytical results for $d=2$ and $d=10$ while the symbols are direct numerical simulations of the auction for $d=2$ ($\diamond$) and $d=10$ ($\bullet$).}
\label{fig:P(x)}
\end{figure}

By definition, $0\le f(x)\le 1$, so that $f (x)^d$ can be neglected compared to $f (x)$ when $d$ is large. This can be verified graphically, by using Eq. (\ref{eq:f(x)}) to express $x$ as a function of $f$. We can then plot $x$ as a function of $f$ and exchange the axes to obtain graphically $f$ as a function $x$. Fig. \ref{fig:f(x)} present $f (x)$ for $d=2$, 3, 4 and 10, from bottom to top. It is easy to appreciate that $\lim_{d\rightarrow\infty} f (x) = 1-x$. In this case, $P (x)$ is a flat distribution. Note also that $df/dx = - P (x)$. Differentiating Eq. (\ref{eq:f(x)}) with respect to $x$, we can express $f$ as a function of $df/dx$ and from Eq. (\ref{eq:f(x)}), $x$ as a function of $df/dx$. Finally, this allows us to draw the graphic of $P (x)$. In Fig. \ref{fig:P(x)}, we show $P (x)$ for $d=2$, 3, 4 and 10, from top to bottom on the left of the figure. Note that we can show that $P(x) \ge 1-1/d$, and that it achieves this minimum value at $x=1$. The curves in Fig. \ref{fig:P(x)} don't all cross at the same point. 

As can be appreciated in Fig. \ref{fig:P(x)}, $P (x)$ is a power-law from $x=0$ up to a critical value $x_c$, where it saturates and becomes a flat distribution. For $P (x) > 1$, the expression for $x (P)$ can be expanded in a series of $1/P$, giving

\begin{equation}
x (P) = \frac{d}{2(d-1)^4} \frac{1}{P^2}+ O \left(\frac{1}{P^3}\right).
\end{equation}
Hence, $P (x) \sim (d^3 x)^{-1/2}$ for $x< x_c$. A higher bound for $x_c$ can be found by setting $P (x_c) = 1$, which gives 

\begin{equation}
x_c \le 1 - \frac{d^{(d-2)/(d-1)}}{d-1} + \frac{d^{-d/(d-1)}}{d-1}.
\end{equation}
In the inset of Fig. \ref{fig:P(x)}, we compare the analytical solution to numerical simulations of the auction for $d=2$ and $d=10$ for $x> 0.1$. The agreement is good.

We conclude that in a strongly competitive environment, where only one player can win, the behaviour of the bid distribution is similar to the one obtained when only two players are competing, at least where the low bid values are concerned. Of course, the lowest bids are of particular interest because deals are usually made at these values.

As in the previous models, it is interesting to consider the distribution of prices, $Z (x)$, corresponding to the probability that the deal will be concluded at a price $x$. From Eq. (\ref{eq:f(x)}), $Z (x)$ is equal to

\begin{equation}
Z (x) = d ( P (x) -1 ) +1. 
\end{equation}
The determination of $Z (x)$ is dependent on the determination of $P (x)$ but, fortunately, even if we do not have any explicit solution for $P (x)$, it is still possible to obtain an expression for $M_1 (d)$, the first moment of $Z (x)$. This first moment, which corresponds to the average price, is calculated by changing $x$ to $f$ in Eq. (\ref{eq:definition of the moments}) using Eq. (\ref{eq:f(x)}). $M_1 (d)$ is equal to 

\begin{equation}
M_1 (d) = \frac{d}{2(d+1)}.
\end{equation}
This result is counterintuitive, as it predicts that when the competition is stronger, with more players, the average price increases. In fact, it shows that when there are a lot of players around, the probability of winning is very small. Hence, players keep on trying to improve and they do not keep any memory of past bids. They are adapting too fast for the game. This could have been anticipated from Fig. \ref{fig:P(x)}, where one sees that the distribution flattens as $d$ increases. Comparing this result with the similar result from the previous $d$-player model, one sees that real life corresponds rather to a game where only the worst player adapts than a game where everyone trys to be the best. 
\section{Market makers}
\label{sec:market makers}

The previous models of auctions are interesting mechanisms to generate an ask price. A buyer solicits several sellers, compares their prices and takes the lowest one available. On the contrary, the generation of a bid price is characterized by a seller considering several buyers. He selects the one offering the highest price. All the results obtained for ask prices $x$ in the previous sections can be transposed to bid prices by changing $x$ to $1-x$.

Most financial exchanges use market makers to add liquidity to the market \cite{hull}. A market maker is a person that will quote both an ask and a bid prices whenever asked to do so. The bid price is the price he is prepared to pay for the asset and the ask price is the price he is prepared to sell the asset at. When solicited, market makers have to give both ask and bid prices because they do not know whether the trader wants to buy or sell the asset. The existence of market makers allows traders to place buy and sell orders whenever they want, without having to wait for somebody else to match their order. This is known as non-synchronical trading. To cover themselves against the risks of possessing unwanted stocks, the ask price proposed by market makers is higher than the bid price. The difference or \emph{spread}, is their risk insurance and their margin for profit. Usually, the exchange regulatory body sets upper limits for spreads. There are of course several market makers on any exchange and they try to quote the lowest ask and highest bid prices, to attract as many traders as possible. However, they cannot afford to be excessively exposed to market risks and have to maintain a minimum spread.

As a more realistic model with direct application to market mechanisms, we consider a mixed auction, where players are market makers. At each time step, the players are required to give a bid and an ask price. Hence, each player has two probability distributions at their disposal. To avoid arbitrage opportunities, each player has to evaluate what the others are likely to propose, such that all ask prices are higher than all bid prices. In practice, the only reference for a market maker is the history of the prices. We assume that a player never proposes an ask price that is lower than a previous winning bid price, thinking that this bid price is likely to be proposed again. Similarly, no market maker will ever propose a bid price that is higher than a previous winning ask price. 

The model works as follows, restricting our attention to a two player game. As we only consider similar initial conditions for both players, we assume that they are using the same distributions. The two players $P$ and $Q$ draw at each time step an ask and a bid price from the same probability distributions $R_a (x,t)$ and $R_b (x,t)$, respectively. The subscript $a$ refers to ask prices and $b$ to bid prices. These prices have to be such that the ask prices are both larger than the higher bid price proposed in the last $h$ time steps, $M_b$, and the bid prices are both smaller than the lower ask price proposed in the last $h$ time steps, $m_a$. $h$ represents the size of the history of the system. We set $M_b = 0$ and $m_a = 1$ at the beginning of the simulations. In case a player selects a bid price higher than $m_a$, he draws a new bid price from a uniform distribution between 0 and $m_a$. Similarly, a player selecting an ask price lower than $M_b$ draws a new ask price from a uniform distribution between $M_b$ and 1. When the trader is a buyer, the market maker with the lowest ask price gets the deal, while for a seller, the market maker with the highest bid price gets the deal. We call $p$ the probability that the trader is a seller. The market maker that does not get the deal discards his proposal and draws a new one, a new ask price if the trader wanted to buy, a new bid price otherwise. New ask prices are drawn from a uniform distribution between $M_b$ and 1, while new bid prices are drawn from a uniform distribution between 0 and $m_a$. We pay no attention to spread requirement.

As at each time step $M_b$ or $m_a$ can be updated, but not both of them simultaneously, we always have $M_b < m_a$. If the trader is a seller, $M_b$ either does not change or increases to a value lower than $m_a$. For a buyer, $m_a$ does not change or decreases to a value higher than $M_b$. Hence, in the limit $h\rightarrow \infty$, $M_b$ and $m_a$ converge to the same value $M$ and stay fixed.  We call this value $M$ the market price. It changes from one simulation to the next, with an average value of $\overline{M} = p$ over several simulations. When the history $h$ is relaxed to a finite value, the market price converges towards $M=p$, for every value of $h$, and oscillates around this value, the larger $h$, the smaller the oscillations.    

The probability distributions of the bid and ask price follow differential equations that depend on $M_b$ and $m_a$. 
For ask prices less than $M_b$, we have

\begin{equation}
\frac{\partial R_a (x,t)}{\partial t} = - (1-p) R_a (x,t)
\label{eq:disparition of R_a}
\end{equation}
and $R_b (x,t)$ follows a similar equation for $x > m_a$. The $1-p$ factor gives the probability that an ask price is required.
The previous equation shows that $R_a (x)$, the stationary limit of $R_a (x,t)$, is zero for $x < M_b$. 
Similarly, $R_b (x)$ is zero for $x > m_a$. 
In reality, both $M_b$ and $m_a$ are functions of time for finite $h$ and the distributions are non-zero on a small interval around $p$. 
However, as already mentioned, $M_b$ and $m_a$ oscillate around a fixed value in the stationary state, so that for the sake of simplicity, we assume that $M_b = m_a = p$, independent of time. The effect of boundary fluctuations is addressed in the next section.
Within this framework, the distribution of ask prices for prices higher than $M_b$ is the solution to 

\begin{equation} 
R_a (x) \int_{p}^x R_a (y) dy = \frac{1}{1-p} \int_{p}^1 R_a (y) \int_{p}^y R_a (z) dz dy.
\end{equation}
Note that we have dropped the time dependence as we only consider the stationary limit. 
This equation is similar to Eq. (\ref{eq:evolution p,2 players}) in the stationary limit, with $p$ as the lower limit instead of 0 and a symmetric condition on the two players. 
As $R_a (x) = 0$ for $x\le p$, it is not necessary to change the lower limits, but we did it for clarity.
The $1/(1-p)$ factor is necessary to allow a proper normalisation of the distribution. 
This arises because the new ask prices are chosen in $(p,1)$, not in $(0,1)$.
Introducing

\begin{equation}
F_{p} (x,t) = \int_{p}^x R_a (y,t) dy,
\end{equation}
we obtain that $F_{p} (x) = \sqrt{(x-p)/(1-p)}$ and $R_a (x) = 1/(2\sqrt{(x-p)(1-p)})$. A similar calculation gives $R_b (x) = 1/(2\sqrt{p(p - x)})$. The solution for the auction of Sec. \ref{sec:the two player game} is obtained for $p=0$, as expected.

In our market maker model, the relative volume of buy and sell orders is controlled by the probability $p$. As explained, the market price $M$ settles close to $M = p$ when the history $h$ is finite. One could wonder to this aspect of the model: when the number of sell orders increases, the model predicts a price increase, while it is well-known than an increase in the supply makes the price go down. We should however stress that $M$ does not represent, in itself, a market price, but that we use this name to simplify the explanations in this section. As stressed in the introduction, $M$ is a measure of the profit made by a market maker whenever a sale is agreed. For $p$ close to 1, market makers are very rarely concluding sales so that they have to make a large profit from each possible sale. They can make a smaller profit from purchases, because the number of occasions to conclude such deals are more aboundant. A more complex model would not only consider $M$ as a profit but as the price of the asset itself. In this case, it should incorporate the fact that market makers are not only competing to increase their number of deals. They have to balance the number of buy and sell orders if they don't want to artificially sustain the price by accepting all sell orders, for instance. As soon as they try to sell, the price will fall quickly, with nobody to match their sale. Hence, the last model has the shortcoming that it does not address the dynamics of matching the orders. As our main interest is on the profit made over the market price, this is not really an issue here. Interesting models where buy and sell orders are matched can be found in \cite{bak96,eliezer98}. 

The previous model can easily be generalised to $d$ market makers and in this case, the results of the previous section can be adapted as we did for the two player auction. To give an idea of a realistic value for $d$, the number of market makers per security varies from a minimum of 2 to a maximum of 68 on the Nasdaq \cite{wahal97}, while George and Longstaff witnessed around 300 market makers among 400 S\&P 100 index option traders \cite{george}. As in the previous sections, the model can also be generalised to cope with prices chosen from a distribution $\omega (x)$ instead of a uniform distribution.
\section{Price volatility as a measure of risk}
\label{sec:price volatility as a measure of risk}

\begin{figure}
\centerline{\psfig{file=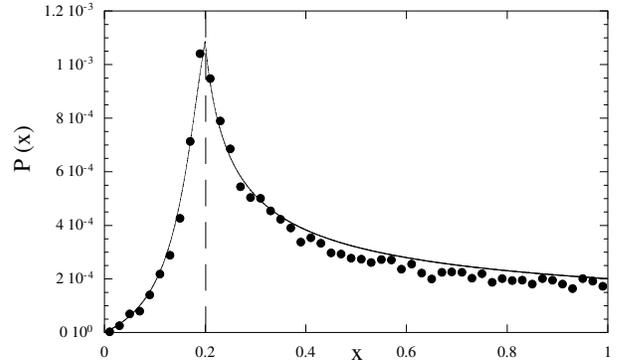,width=8.5cm}}
\caption{Probability distibution for the bids of the players. The points correspond to a direct simulation of the model, while the continuous line is a numerical solution of Eqs. (\ref{eq:solution for 0-x-delta}) and (\ref{eq:solution for delta-x-1}). The dashed lines indicate the level of uncertainty, fixed to $\Delta=0.2$ during the simulation.}
\label{fig:P (x)-price fluctuations}
\end{figure}

Up to this point, the only uncertainty facing the players has been the decision of the other players. In reality, a major source of uncertainty can be found in price fluctuations. This corresponds in our framework to variations in the mininum profit necessary to hedge against market fluctuations. We consider a simple auction model where two players $P$ and $Q$ propose ask prices drawn from the range $(M , 1)$, with $M$ chosen from a uniform distribution in $(0, \Delta )$ at each time step. As in the model of Sec. \ref{sec:the two player game}, $P$ and $Q$ are given probability distributions, $P (x, t)$ and $Q (x,t)$ respectively, to choose their bids. Whenever a chosen bid is less than $M$, it is discarded and another bid chosen at random from the range $(M,1)$ is proposed. The player with the lowest bid gets the deal. As we consider similar initial conditions for both players, the probability distribution $P (x, t)$ follows 

\begin{equation}
\frac{\partial P (x,t)}{\partial t} = - P (x,t)
\end{equation}
for $x\le M$ and 

\begin{equation}
\frac{\partial P (x,t)}{\partial t} = - P (x) \int_0^x P (y,t) dy + \frac{\alpha}{1-M}
\end{equation}
for $x\ge M$. We have defined $\alpha$ as the probability that $Q$ wins, as in Eq. (\ref{eq:definition of alpha}). Considering stationary solutions and from our choice for the dynamics of $M$, $P (x)$ is the solution to

\begin{equation}
P (x) \left( 1- \frac{x}{\Delta} + \frac{x}{\Delta} \int_0^x P (y) dy \right) = -\frac{\alpha}{\Delta} \ln (1 - x)
\label{eq:solution for 0-x-delta}
\end{equation}
for $0 \le x\le \Delta$ and
\begin{equation}
P (x) \int_0^x P (y) dy = - \frac{\alpha}{\Delta} \ln (1 - \Delta)
\label{eq:solution for delta-x-1}
\end{equation}
for $\Delta \le x\le 1$. The exact solution to the second equation is

\begin{equation}
P (x) = N_0 \left( \frac{P^2 (\Delta) \alpha \ln (1 - \Delta)}{\alpha \ln (1-\Delta) + 2 P^2 (\Delta) \Delta (\Delta -x)}\right)^{1/2}
\end{equation}
for $\Delta \le x\le 1$, where $N_0$ is a normalisation coefficient. 
The distribution is a power-law in this range, with the same exponent as in the two player auction of Sec. \ref{sec:the two player game}. We could not solve the first equation, but its numerical solution can be compared with direct simulations of the model. The results are presented in Fig. \ref{fig:P (x)-price fluctuations} and as can be seen, Eq. (\ref{eq:solution for 0-x-delta}) compares very well with the model.

The previous analysis leads us to the conclusion that the distributions obtained in the framework of simple auction models are robust outside the range of price fluctuations. However, agents should be reluctant to propose prices inside this range, in agreement with usual pricing models that take the volatility as a proxy to investment risk \cite{hull}. The simple model presented here can of course be easily generalized to deal with more realistic price fluctuations. Even if we think that the present conclusion should remain applicable for most situations, it is however important to stress that price fluctuations are known to be non-Gaussian \cite{bouchaud}. Large fluctuations are not so rare and that has a major impact on the value of $M$. Hence, in period of quiescence, the conclusion of this section should apply, while for more agitated markets, what we call the range $M$ of the fluctuations could force agents to propose prices well inside the range of fluctuations.
\section{Implementing a market structure}
\label{sec:implement a market structure}

In the previous models, sellers are competing in an abstract infinite dimensional space, where every trader is identical, apart possibly from initial conditions. However, lots of trades rely on a strict market structure, where buyers are interacting with only a restricted set of sellers. To investigate the effect of space on sellers price distribution, we consider a variation of the $d$ player game of Sec. \ref{sec:the d player game} where players are nodes of a network, competing with their first neighbours. It should be obvious that an important quantity in such a framework is the connectivity distribution, that is, the number of sellers you are competing with. But who you are competing with is also of major importance. For a regular network with $d-1$ neighbours for each site, the $d$ player game is recovered. More interesting is the case where a site has $k$ neighbours with a probability $c_k$. As for the $d$ player game of Sec. \ref{sec:the d player game}, two extreme situations can be considered; either a player is happy unless he gets no deal, or a player is happy only if he gets all possible deals. Based on our conclusions of Sec. \ref{sec:the d player game}, we only consider the former. 

There is one customer and one player at each node of the network. At each time step, every player proposes a price drawn at random from his personal bid distribution. A customer at one node buys from the cheapest price among the price proposed by the player at his node and the prices proposed by the players of the neighbouring nodes. Hence, a player located at a side with $k$ neighbours can get from 0 to $k+1$ customers in every round of bidding. As long as a player gets one customer, he does nothing, while a player with no customer discards the price he proposed and draws a new price at random from a uniform distribution. The bid distribution of a player $P$ with $k$ first neighbours, $Q_i$, with $i = 1$, ..., $k$, evolves according to

\begin{equation}
\frac{\partial P (x,t)}{\partial t} = - P (x,t) \prod_{i=1}^k \int_0^x Q_i (y,t) dy + 1 - \alpha_P
\end{equation} 
where $\alpha_P$ is the probability that $P$ wins and $Q_i (x,t)$ the bid distributions of the neighbours of $P$. Of course, these distributions follow similar evolution equations involving their own neighbours. With the chosen updating rule, winning is synonymous with getting at least one deal.

We are not able to solve the previous set of equations in the general situation, but it is tempting to assume that $Q_i (x) = (1-\alpha_i) x^{-\alpha_i}$ in the stationary state, by analogy with the previous models. In this case, it is easy to show that the condition imposed on the exponent at one site is that the sum of this exponent and the exponents of all neighbouring sites is equal to the number of neighbours. However, this solution is not compatible with the condition $\alpha_P \in (0,1)$, at least for some special situations, hinting that this works only for special cases, when all players have the same number of neighbours for instance. To show that the previous assumption does not capture the complete picture, we consider the stationary limit of a very simple network of $k+1$ nodes. $k$ of these nodes have only one link pointing to the central $k+1^{\hbox{\small th}}$ node. This corresponds to one central seller $P$ trying to compete with $k$ local sellers, $Q_i$, with $i=1$, ..., $k$. By symmetry, all local sellers should have the same distribution $Q_i (x) \equiv Q (x)$. One could think of the central node as a supermarket and all the neighbouring nodes as small differenciated shops. As the small shops do not sell similar goods, they do not compete with each other, while the supermarket is competing on all goods. The fact that we take only one price for all goods is justified by the fact that the different goods prices are correlated, being all sold by the supermarket. Alternatively, some particular geographical situation could make going to other shops uninteresting, like restricted parking places, while the supermarket could provide easy access. In this case, we obtain in the stationary limit that

\begin{equation}
P (x) \left( \int_0^x Q (y) dy \right)^k = 1 - \alpha_P
\end{equation} 
and

\begin{equation}
Q (x) \int_0^x P (y) dy = 1 - \alpha_Q.
\end{equation} 
Introducing $G (x) = \int_0^x Q (y) dy$, we can show that

\begin{equation}
(\alpha_Q - 1) G (x)^k \frac{\partial^2 G (x)}{\partial x^2} = (1 - \alpha_P) \left( \frac{\partial G (x)}{\partial x}\right)^2.
\end{equation}
The previous equation can be solved to obtain 

\begin{equation}
Ax + B = \int dG\ \exp \left( \frac{(1 - \alpha_P)G^{1-k}}{(1 - \alpha_Q)(1-k)} \right)
\end{equation}
where $A$ and $B$ are integration constants. Using $G (0) = 0$ and $G (1) = 1$, we obtain the implicit solution

\begin{equation}
1 - x = \frac{\Theta (G^{1-k}) }{\Theta (\infty)}
\end{equation}
where we defined

\begin{equation}
\Theta (v) = \int_1^v u^{\frac{k}{1-k}} \exp \left(\frac{(1 - \alpha_P) u}{(1 - \alpha_Q) (1-k)}\right) du.
\end{equation}
This shows that for $x$ close to 1, $Q(x)$ is uniform, while $P (x) \sim x^{-k}$. For $x$ close to 0, the leading term for $Q (x)$ is $1/x$, with important logarithmic corrections, and it can be written

\begin{equation}
Q (x) \sim \frac{1}{x \left(\ln (x/\gamma ) \right)^{k/(k-1)}} 
\end{equation}
for some function $\gamma$ that depends on $\alpha_P$, $\alpha_Q$ and $k$. In this limit, $P (x) \sim x Q(x)$.
\begin{figure}
\centerline{\psfig{file=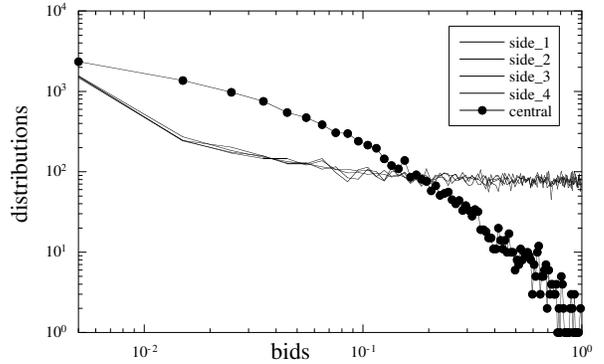,width=8.5cm}}
\caption{Probability distibutions for the bids of the players for one central seller and 4 side sellers competing with him. The continuous lines are for the side sellers, while the continuous line with dots ($\bullet$) is the distribution of the central seller.}
\label{fig:P (x)-price for one central}
\end{figure}
In Fig. \ref{fig:P (x)-price for one central}, we present the results of a simulation where one central seller competes with 4 side sellers, that do not compete with each other. Each seller had $10^4$ different prices at their disposal and they played $10^7$ rounds of bidding. For the particular simulation presented, $1 - \alpha_P \approx 0.03$ and $1 - \alpha_Q \approx 0.7$. As mentionned earlier, the different probabilities do not have to sum up to 1, as they don't refer to exclusive events. As can be appreciated in Fig. \ref{fig:P (x)-price for one central}, for $x$ close to 1, $Q(x)$ becomes uniform, while $P (x)$ can arguably said to converge towards a power law. We could not check numerically the value of the exponent of this power law because it extends over less than a decade.

In Fig. \ref{fig:P (x)-price for one central}, it is apparent that the distribution for the side players is uniform from 0.1 to 1, which signals a bad adaptation. This impression is justified by the fact that they do not win very often. In simple terms, the central player benefits from having to compete with several players. This is to be expected, as the central player wins if he is not the worst player out of $k$, while the side players have to be the best of two not to lose. We arrive at the interesting conclusion that, as in Sec. \ref{sec:the d player game}, it pays not to change prices too often. In fact, new prices being choosen from $(0,1)$, they are unlikely to be competitive. Assuming that we can extend this conclusion to more general networks, we expect to find that sites with a larger connectivity, surrounded by sites with smaller connectivity, are winning more often. This conclusion has some echoes in real life, where supermarkets benefit from attracting a wider range of customers than small shops.
\section{Conclusions}
\label{sec:conclusions}

We have introduced simple models for tendering processes in an attempt to model the dynamics of intermediaries trying to make a profit from the competitive sale of a commodity or a service. Starting with a simple 2-player game, we extended our model to a $d$-player game, considered the problem of market making, investigated the effect of price fluctuations and implemented a strict market structure. In all these cases, the bid distribution of the players has been our main concern and we showed that it can strongly depend on the system details. Nevertheless, two generic features could be seen in all these models. First, unless there is a strict market structure to differentiate the players, they tend to become identical in the long time limit. This stationary state corresponds to a maximum profit state for the whole system. Hence, cooperation has appeared in a system made up of selfish individuals, a property reminiscent of the Minority Game \cite{challet97}. Second, we showed that players generally benefit from waiting longer before updating their beliefs. Players updating their bid distribution too quickly are not able to discern between a trend and a fluctuation. This was also spotted in an evolutionary variant of the Minority Game, where it was shown that players prefer to keep on playing with only one strategy  \cite{johnson98-2,dhulst99}. We can compare this particularity to models of growing boundaries, where noise can be drastically reduced by only allowing sites that have been selected a given number of times to grow \cite{barabasi}. Similarly, we can improve the adaptation process by updating bids only if they have lost a given number of times.

Considering each variant of the model separately, we can refine our conclusions. For the two-player game, the stationary bid distribution is a power law and this result can be extended to a $d$-player game. However, this result is dependent on the type of adaptation process chosen, and we showed that it would not be the solution of a simple generalisation. By considering the average price generated by the model, which corresponds to a player's expected profit, we showed that a model with only the worst player adapting compares better with reality than a model in which all the non-winning players adapt. This suggests that in real life, unless you are the worst, you still make a profit from business. Considering a simple generalisation to mimic market making, the previous results have been extended. The major difference is that the reference price, corresponding to 0 in the first models, is not fixed now. This leads us to consider the effect of price uncertainty over the bid distribution. We showed that outside the range of the fluctuations, the preliminary result obtained in the simple models can be extended. So, in a quiescent market, one expects our result to apply, while for unsettled markets like markets from emerging countries, fluctuations are so important that they should strongly affect the bid distribution of the participants. Finally, putting the players on a network allowed us to generate heterogeneous players. We showed that a player's bid distribution is a function of his neighbours. From a simple example, we concluded that players with a high connectivity connected with players of low connectivity are optimal, in the sense that they should get most of the deals.

\end{document}